\documentclass[aps,pre,twocolumn,letterpaper,superscriptaddress]{revtex4}
\usepackage{graphicx}
\usepackage{epstopdf}
\usepackage{amsmath,color}
\usepackage{amssymb}
\usepackage{textcomp}
\usepackage{url}
\usepackage{hyperref}

\begin{document}

\author{Reza Sepehrinia}\email{sepehrinia@ut.ac.ir}\affiliation{Department of Physics, University of Tehran, P. O. Box 14395-547, Tehran, Iran}\author{Abbas Ali Saberi}\email{ab.saberi@ut.ac.ir (corresponding author)}
\affiliation{Department of Physics, University of Tehran, P. O. Box 14395-547, Tehran, Iran}
\affiliation{Institut f\"ur Theoretische
  Physik, Universit\"at zu K\"oln, Z\"ulpicher Str. 77, 50937 K\"oln,
  Germany}\author{Hor Dashti-Naserabadi}
\affiliation{School of Physics, Korea Institute for Advanced Study, Seoul 02455, South Korea}

\title{Random Walks on Intersecting Geometries}

\begin{abstract}
We present an analytical approach to study simple symmetric random walks (RWs) on a crossing geometry consisting of a plane square lattice crossed by $n_l$ number of lines that all meet each other at a single point (the origin) on the plane. The probability density to find the walker at a given distance from the origin either in a line or in the plane geometry is exactly calculated as a function of time $t$. We find that the large time asymptotic behavior of the walker for any arbitrary number $n_l$ of lines is eventually governed by the plane geometry after a crossover time approximately given by $t_c\propto n_l^2$.
We show that this competition can be changed in favor of the line geometry by switching on an arbitrarily small perturbation of a drift term in which even a weak biased walk is able to drain the whole probability density into the line at long time limit. We also present the results of our extensive simulations of the model which perfectly support our analytical predictions. Our method can, however, be simply extended to other crossing geometries with a single common point.
\end{abstract}

\maketitle

\section{Introduction}

Random walks (RWs) are ubiquitous models of stochastic processes playing an essential role in many challenging problems in probability and statistical physics \cite{Feller1968, Feller1971, Hughes1995, Masuda2017}. For random walks, the probability density $\rho(r,t)$ to
find a walk at time $t$ at a site with distance $r$ from
its origin, obeys the scaling collapse \cite{Havlin1987}
\begin{eqnarray}\label{rho}
   \rho(r,t)\sim t^{-d_f/d_w}f(r/t^{1/d_w}),
\end{eqnarray}
with the scaling variable $r/t^{1/d_w}$,  where $d_f$ is the dimension of the underlying (possibly fractal) network. On a lattice with translational invariance symmetry in any spatial dimension $d$ ($=d_f$), it
has been shown that the walk is always purely diffusive, i.e.,
$d_w = 2$, with a Gaussian scaling function $f$, which has been
the content of many classic textbooks on random walks
and diffusion \cite{Weiss1994, Feller1968}. The scaling relation (\ref{rho}) still remains valid when translational invariance is blurred in certain ways
or the network is fractal (i.e., for non-integer $d_f$). However, anomalous diffusion with $d_w\ne 2$ may arise in various transport processes \cite{Havlin1987, Bouchaud1990, Redner2001}.\\ Using equation (\ref{rho}), it is now straightforward to conceive the P\'{o}lya's recurrence theorem \cite{Polya1921} that a simple symmetric RWs on $\mathbb Z^d$ lattice
is recurrent in $d\le2$, but transient in $d\ge3$. Also widely known is that the
transition between recurrence and transience occurs precisely at $d=2$, rather than at
some fractal dimension $2<d_f<3$. In this sense, $d=2$ is the "critical dimension"
for intersection of a two-dimensional set (i.e., the path of RWs) and a
zero-dimensional set (the origin).
Moreover, it is known \cite{Lawler2010} that the scaling limit of the simple RWs in $d$ dimensions converges to the $d$-dimensional standard Brownian motion which has a certain invariance under conformal maps in two-dimensions. The conformal invariance in $d=2$ then provides a powerful tool to exactly determine the values of the involved exponents \cite{Lawler2010}.

\begin{figure}
	\centering
	\includegraphics[width=0.48\textwidth]{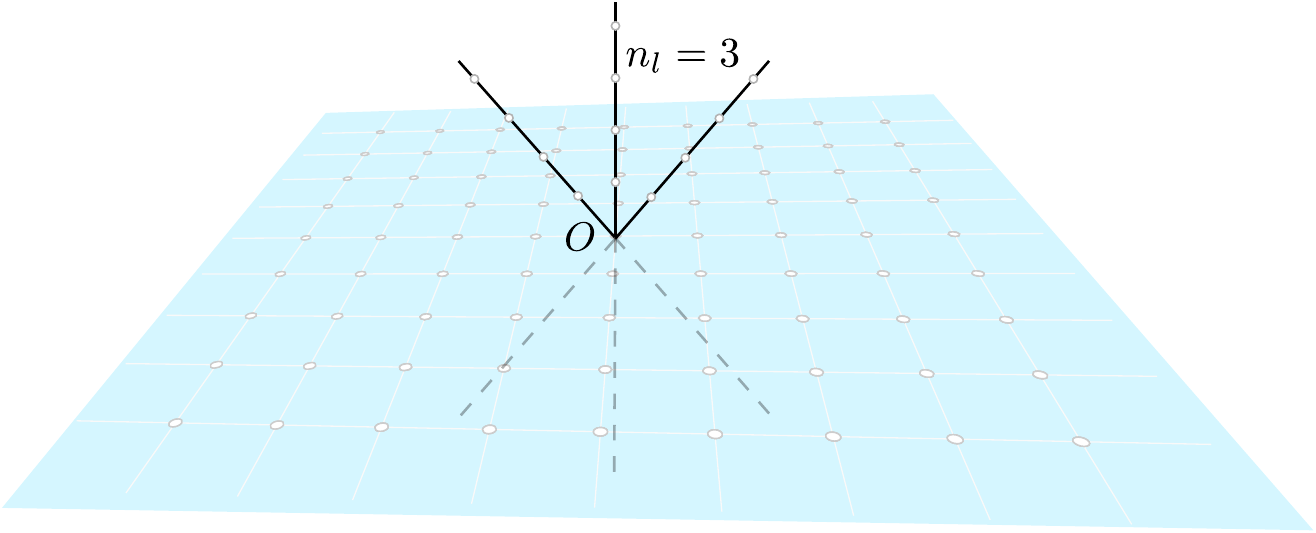}
	\caption{Illustration of the combined geometry in our model composed of an infinite lattice plane and $n_l$ number of crossing lattice lines that all share a single common point---the origin.}
	\label{fig:plane_lines}
\end{figure}

Here we consider the RWs on a mixed geometry consisting of an integer lattice $\mathbb Z^2$ which is crossed by $n_l$ number of lattices $\mathbb Z$ that all share a single common point---the origin (see Fig. \ref{fig:plane_lines}). The RWs initiate from origin $\textbf{x}=0$ at time $t_0=0$ and we ask if the  statistics of the walks at time $t$ obeys the scaling form (\ref{rho}).

Let us list the main results of this paper. (1) A competing behavior is observed in which at early times the crossing line geometries are dominant and become less effective in time until a crossover time $t_c$, after which the plane geometry governs the statistics of the RWs.

(2) The probability density to find the random walker at the origin behaves like $\rho(0,t)\sim t^{-\alpha}$ with $1/2\le \alpha\le 1$ spanning the crossover behavior from early time $t\ll t_c$ with $\alpha=1/2$ for the line geometry ($d=1$ and $d_w=2$ in Eq. (\ref{rho})), to long-time limit $t\gg t_c$ with $\alpha=1$ for the plane geometry ($d=2$ and $d_w=2$ in Eq. (\ref{rho})). Therefore, the symmetric RWs is always recurrent even for arbitrarily large number of crossing lines ($n_l\gg 1$) which may assign a larger effective dimensionality ($d>2$) to the whole geometry.

(3) We find both analytically and numerically that the crossover time $t_c$ grows with the number of crossing lines $n_l$ with the approximate power-law relation $t_c\sim n_l^2$.

(4) Our analytical prediction for the mean squared displacement of the RWs on a crossing line at long-time limit $t\gtrapprox t_c\sim n_l^2$, gives $\langle z_l^2\rangle\sim \sqrt{2/\pi^3}  n_l \sqrt{t} \log t$, which is well supported by our results obtained from numerical simulations of the model.

(5) The probability to find the RWs at a point $\textbf{r}_p\equiv (x,y)$ on the plane or at a point $z_l$ on a line at time $t$ is provided by the generating function given in Eq. (\ref{Px}).

The rest of this paper is structured as follows. In Section \ref{formulation}, we will present a general formulation of our model for general combined lattices
and briefly discuss its long-time behavior. In Section \ref{example} we will study an interesting nontrivial
example of the model composed of a lattice plane and a chain which share a single point.
Section \ref{biased} will discuss the asymmetric RWs on the chain competing with a lattice plane and discuss its
asymptotic behavior. In Section \ref{simul} we will present the results of our numerical simulations for a plane lattice crossed by $n_l$ number of chains which show perfect agreement with our analytical results. Finally, the last section concludes our discussion.

\begin{figure}
	\centering
	\includegraphics[width=0.47\textwidth]{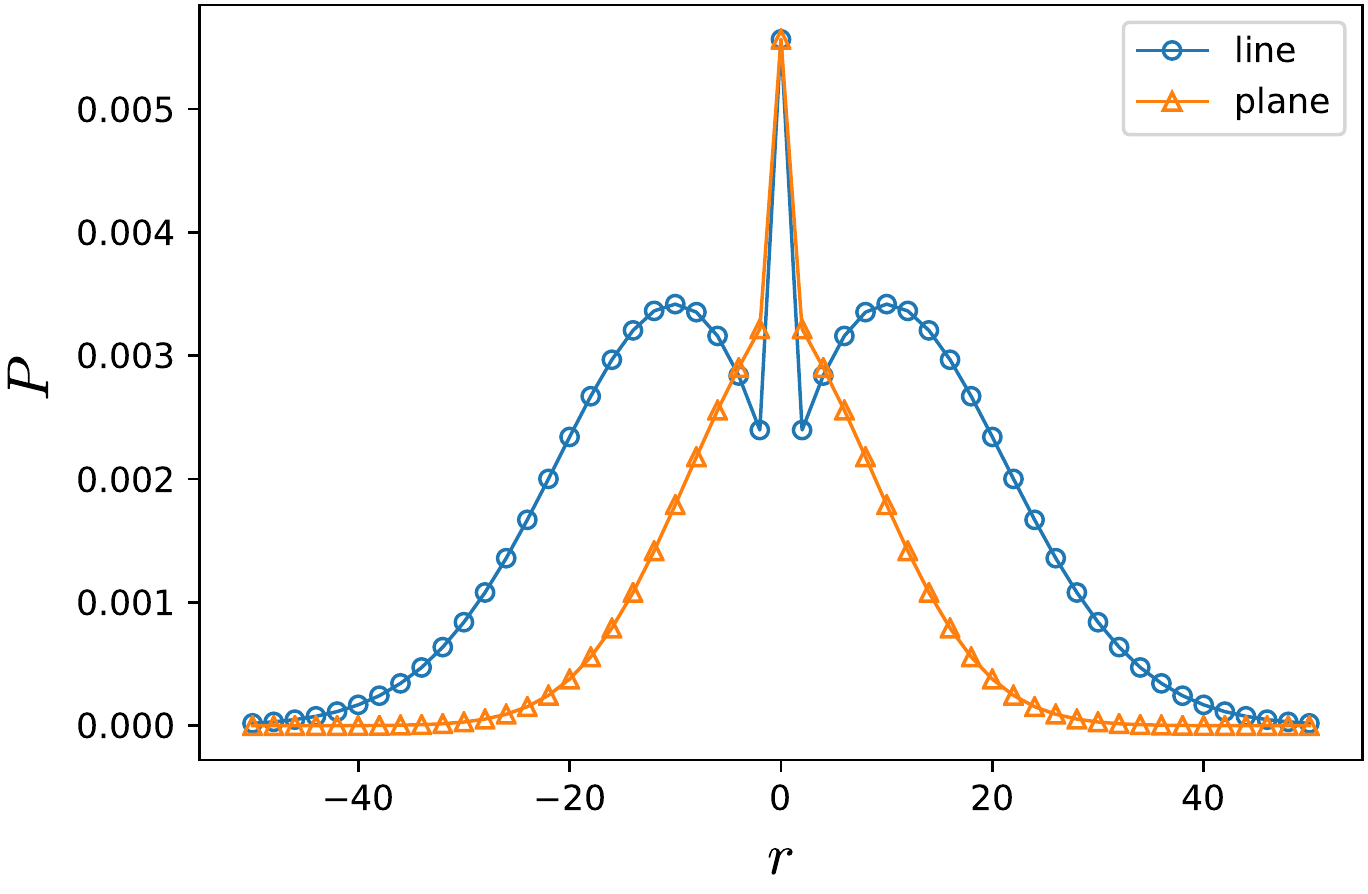}
	\caption{(color online) Probability of finding the RWs at distance $r$ from the origin either on a line (circles) or on the plane (triangles) at time $t=200$ for $n_l=1$ obtained from numerical inverse $z$-transform of Eq. (\ref{Px}). Both probability functions get wider in time. }
	\label{fig:P(x)}
\end{figure}

\begin{figure}
	\centering
	\includegraphics[width=0.47\textwidth]{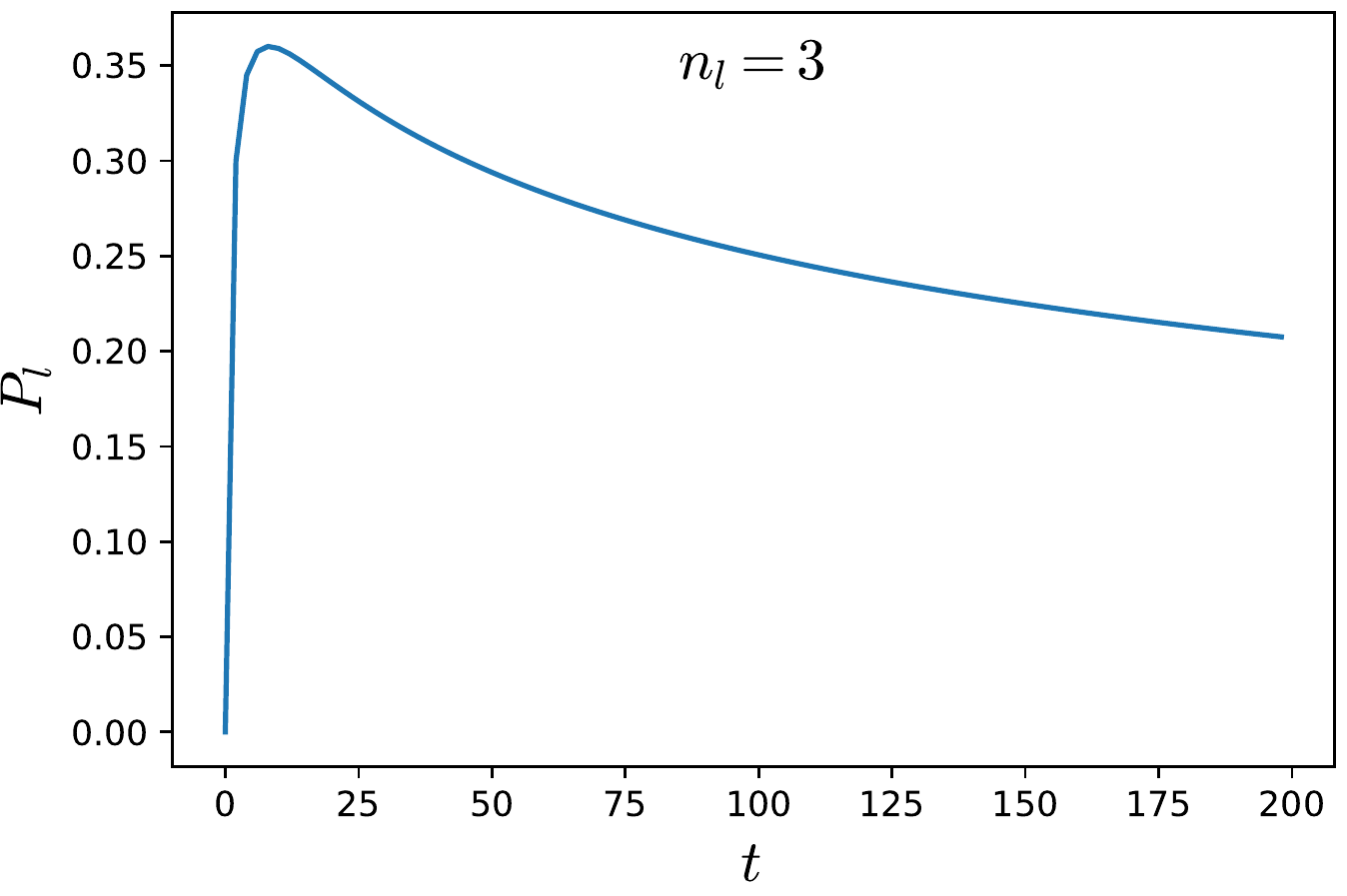}
	\caption{(color online) Total probability for the walker to being on one of the line geometries as a function of time (for even time steps $t=2n$).}
	\label{fig:Pl}
\end{figure}

\section{General Statement and  Formulation of the Problem}\label{formulation}

We consider two general lattices $a$ and $b$, on which the random walk problem is known. We pose the following question: what would be like the statistics of the random walk motion if we connect $a$ and $b$ in a way that they have a single point in common which we call $O$ $^{\text{1}}$ \footnotetext[1]{This problem can well be interpreted as finding the quantum mechanical Green's function of a single particle moving on the underlying lattices.}. Consider a classical symmetric random walk that starts from origin $O$. We would like to determine the probability of finding the walker at a given point (on $a$ or $b$) at time step $n$. The most simple quantity to determine for the combined geometry is the first passage probability through the origin. Let us denote the probability of arriving at $O$ for the first time at the $n$th step by $F_\textbf{0}(n)$. For this quantity the walker is required not to visit the point $O$ until the $n$th step. Therefore once it stepped into $a$ or $b$ right after the first walk, it should remain there and return to $O$ at the $n$th step. Depending on the coordination number at $O$, the only shared point between the two domains $a$ and $b$, at the first step the walker would enter into either $a$ or $b$ with probabilities $p_a$ and $p_b$, respectively, where $p_a+p_b=1$. We now have
\begin{eqnarray}\label{F_0}
F_\textbf{0}(n)=p_a F^a_\textbf{0}(n)+p_b F^b_\textbf{0}(n),
\end{eqnarray}
where $F^\textit{i}_\textbf{0}(n)$ with $i=a, b$ denotes for the same quantity as $F_\textbf{0}(n)$ for either isolated lattice. Using this simple relation, one can immediately obtain the total return probability $R$ to the origin,
\begin{eqnarray}\label{Return}
R=\sum^{\infty}_{n=1}F_\textbf{0}(n)=p_a R^a+ p_b R^b.
\end{eqnarray}
If the random walk is recurrent on each of the two lattices $a$ and $b$, i.e., $R^a=  R^b=1$, then it will be recurrent on the combined geometry too i.e., $R=1$.

With the first passage probability in hand, one can find the site $\textbf{x}$ occupation probability $P_\mathbf{x}(n)$ as well. But let us first consider the case for the origin, i.e., $\textbf{x}=\textbf{0}$, by letting $P_\textbf{0}(n)$ denote the probability of finding the walker at origin at the $n$th step. This can be expressed in terms of the first passage probability through $O$ with the following relation \cite{Hughes1995,Montroll1965}
\begin{eqnarray}\label{P_o}
P_\textbf{0}(n)=\delta_{0n}+\sum_{i=1}^n F_\textbf{0}(i) P_\textbf{0}(n-i),
\end{eqnarray}
in which the summation is assumed to be zero for $n=0$. Using $z$-transform, as a discrete-time equivalent of the Laplace transform, on both sides of the equation (\ref{P_o}) by multiplying both sides by $z^n$ and summing over $n$, one can find a simple equation for the generating function \begin{eqnarray}\label{P(z)}P_\textbf{0}(z)=1+F_\textbf{0}(z)P_\textbf{0}(z),\end{eqnarray} in which we have used  $P_\textbf{0}(z)=\sum_n z^n P_\textbf{0}(n)$ and similar relation for $F_\textbf{0}(z)$. Using equations (\ref{P(z)}) and (\ref{F_0}) gives
\begin{eqnarray}
   P_\textbf{0}(z)&=&\left(1-F_\textbf{0}(z)\right)^{-1}, \nonumber \\
   &=&\left(1-p_a F^a_\textbf{0}(z)-p_b F^b_\textbf{0}(z)\right)^{-1},
\end{eqnarray}
which together with the normalization condition $p_a+p_b=1$, leads to the following result
\begin{eqnarray}\label{P0}
   \frac{1}{P_\textbf{0}(z)}=\frac{p_a}{P^a_\textbf{0}(z)}+\frac{p_b}{P^b_\textbf{0}(z)},
\end{eqnarray}
that is very akin to the reciprocal of the total equivalent resistance of two parallel resistors.

\begin{figure}
	\centering
	\includegraphics[width=0.47\textwidth]{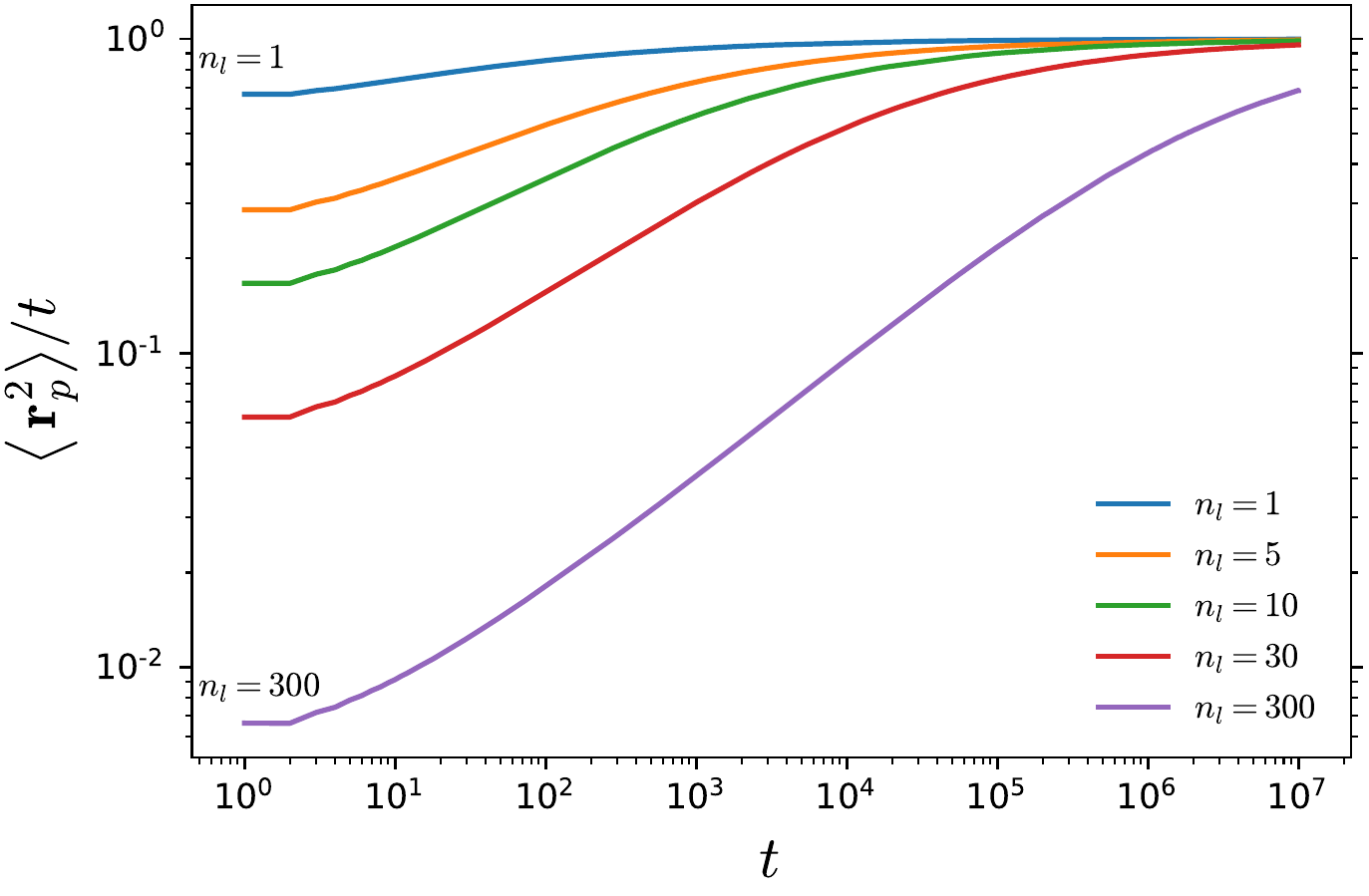}
	\caption{(color online) Mean squared displacement of the RWs over time $t$ on the plane, i.e., $\langle \textbf{r}^2_p\rangle/t$, as function of $t$ in logarithmic scale for different number of crossing lines $n_l=1$ to $300$ from top to bottom. All data converge to the diffusion constant $D_p=1$ on the plane at long-time limit.}
	\label{fig:rp2}
\end{figure}

Now let us calculate the site occupation probability $P_\mathbf{x}(n)$ at a given site $\mathbf{x}$ other than the origin. This quantity can be determined in terms of the solutions in the individual geometries. The probability to arrive at $\mathbf{x}$ at the $n$th step can be considered as the sum of the probability of being at the origin at any earlier time $i<n$ and arriving to the destination without visiting the origin on the remaining time $n-i$. The latter is known as the 'taboo' probability \cite{Hughes1995} denoted by $T_{\textbf{x}}(n-i)$, in which the walker avoids the origin. One can therefore find that $P_\mathbf{x}(n)$ can be cast into the following form \begin{eqnarray}\label{Px}P_{\textbf{x}}(n)=\sum_{i=0}^{n-1} P_{0}(i) T_{\textbf{x}}(n-i),\end{eqnarray} for which $z$-transformation provides $P_\textbf{x}(z)=P_\textbf{0}(z)T_\textbf{x}(z)$. Since the walker has to avoid the origin, it should stay in one of the either $a$ or $b$ geometries during the time interval $(i,n]$. This means one can write
\begin{eqnarray}\label{Px}
   P_{\textbf{x}}(z)=P_\textbf{0}(z)\times\left\{
   \begin{array}{cc}
    p_aT^a_\textbf{x}(z) & \textbf{x}\in a \\
    p_bT^b_\textbf{x}(z) & \textbf{x}\in b
   \end{array}\right.
\end{eqnarray}
For a translationally invariant lattice one can show that for $\textbf{x}\neq 0$, $T^{a,b}_{\textbf{x}}(z)=F^{a,b}_{\textbf{x}}(z)=P^{a,b}_{\textbf{x}}(z)/P^{a,b}_{\textbf{0}}(z)$.
%
%

\begin{figure}
	\centering
	\includegraphics[width=0.47\textwidth]{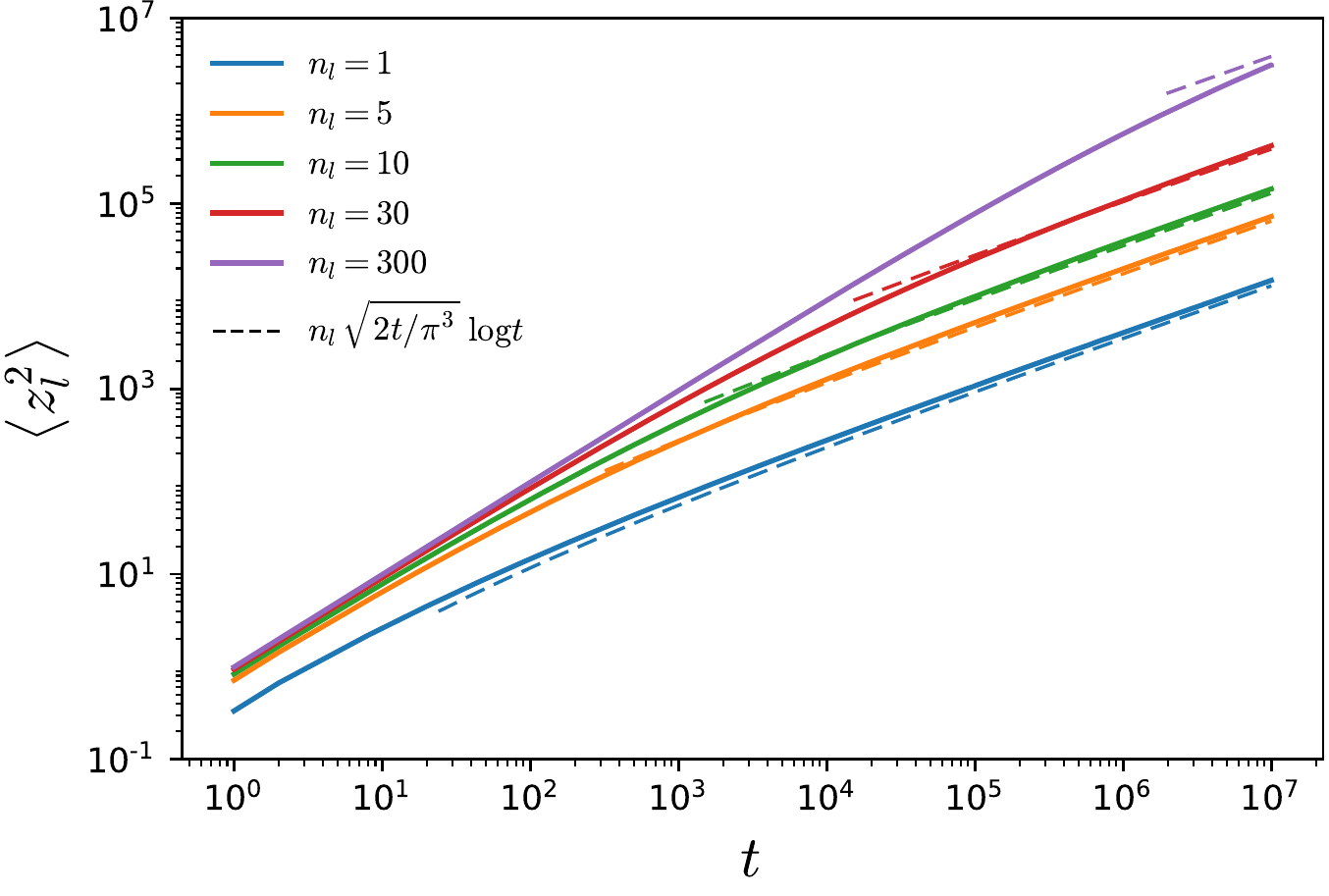}
	\caption{(color online) Mean squared displacement of the RWs on a crossing line, i.e., $\langle z_l^2\rangle$, as function of $t$ in logarithmic scale for different number of crossing lines $n_l=1$ to $300$ from bottom to top. The dashed lines show the comparison with our analytical prediction of the behavior at long-time limit, i.e., $\langle z_l^2\rangle\sim \sqrt{2/\pi^3}  n_l \sqrt{t} \log t$ for $t\gtrapprox t_c\propto n_l^2$---see Fig. (\ref{fig:Prob}).}
	\label{fig:z2}
\end{figure}

\subsection{long-time asymptotics}
In order to determine the probabilities as function of time from their $z$-transforms,
one needs to do inverse $z$-transformation which means to find the coefficients of
Taylor series about $z=0$. It is often of more interest to look at the behavior
in the long-time limit which is encoded in $z\rightarrow 1^-$ limit of the corresponding
$z$-transform. For instance, if the lattice $a$ has the more rapidly decreasing probability
then its $z$-transform is less divergent. As a result, Eq. (\ref{P0}) gives $P_\textbf{0}(z)\overset{z \rightarrow 1^{-}}{\approx} P^a_\textbf{0}(z)/p_a$, meaning that the long-time behavior is governed by the lattice $a$. Then the Tauberian theorem \cite{Hughes1995,Hardy1949} can be used to obtain the asymptotic behavior in time domain.
\section{$1d$ lattice and Square lattice}\label{example}
In this section, we are going to study an interesting nontrivial example of the general formulation presented in the previous section by taking $a$ a square lattice in $x$-$y$ plane and $b$ a one-dimensional (1$d$) lattice which crosses $a$ at a single point---the origin $O$ (Fig. \ref{fig:plane_lines}). The RW problem is exactly solvable on these two lattices with known results \cite{Montroll1965,Hughes1995}
\begin{eqnarray}\label{P0-l}
P^l_{\textbf{0}}(z)&=&\frac{1}{\sqrt{1-z^2}}\overset{z \rightarrow 1^-}{\approx}[2(1-z)]^{-\frac12}\\ \label{P0-p}
P^p_{\textbf{0}}(z)&=&\frac{2}{\pi}K(z^2)\overset{z \rightarrow 1^-}{\approx}\frac{1}{\pi}\log[1/(1-z)],
\end{eqnarray}
where $K(x)$ is the elliptic integral of first kind. \\Long-time $n \rightarrow \infty$ behavior of the occupation probability of the origin is decreasing algebraically with time given by $P^l_{\textbf{0}}(n) \approx\frac{1}{\sqrt{2\pi n}}$ and $P^p_{\textbf{0}}(n) \approx\frac{1}{\pi n}$ on the chain and the square lattice, respectively. Using Eq. (\ref{P0}) with $p_a=\frac{2}{3}$ and $p_b=\frac{1}{3}$ for the combined geometry and noting that the
second term in the right-hand side is dominant in the limit $z\rightarrow 1^-$ we have
$P_{\textbf{0}}(n) \approx \frac{3}{2\pi n}$. Since each lattice is translationally invariant as we mentioned for $\textbf{x}\neq 0$ we have $T^{l,p}_{\textbf{x}}(z)=P^{l,p}_{\textbf{x}}(z)/P^{l,p}_{\textbf{0}}(z)$. Therefore it is enough to know $P^{l,p}_{\textbf{x}}(z)$ in order to calculate the probabilities on other lattice points of combined lattice.
For these two lattices it is also analytically available.
For example $P^l_{\textbf{x}}(z)=(1-z^2)^{-1/2}(1-\sqrt{1-z^2})^{|x|} z^{-|x|}$ and $P^p_{\textbf{x}}(z)$
can be expressed in terms of hypergeometric functions \cite{Ray2014}. Using Eq. \ref{Px} we obtain the site
probability for a given lattice point. A plot of probability as a function of position at a given time is
shown in Fig. \ref{fig:P(x)}  for line and plane. To see how the diffusion takes place on the line
for example, we can calculate the moments of the probability distribution. Total probability of finding
the walker i.e. the zeroth moment on the line, $P^l(n)=\sum_{\textbf{x}\in l} P_\textbf{x}(n)$, is given by
\begin{equation}\label{mu0l}
 P^l(z)=  \frac{1}{3}\frac{2z}{1-z+\sqrt{1-z^2}} P_{\textbf{0}}(z)
\end{equation}
Figure \ref{fig:Pl} shows the plot of this moment as a function of time. As we can see the total probability of finding the walker on the line raises first and then decreases. Using Eq. \ref{mu0l} we can see that the long time behavior of the total probability on the line is like $P^l(n)\sim \frac{1}{\sqrt{n}}\log(n)$. The first moment vanishes because of symmetry under $\textbf{x} \rightarrow -\textbf{x}$. To see how fast the particle diffuses on the line, it is also worth calculating second moment $\langle z_l^2 \rangle=\sum_{\textbf{x}\in l} |\textbf{x}|^2 P_\textbf{x}(n)$. We find
\begin{equation}\label{mu0l}
 \langle z_l^2 \rangle=  \frac{1}{3}\frac{z(1+z)^{1/2}}{(1-z)^{3/2}} P_{\textbf{0}}(z)
\end{equation}
The time dependence at large time is like $\langle z_l^2 \rangle\sim \sqrt{n}\log(n)$.

Now we consider intersection of $n_l$ lines with a plane at the origin. Due to symmetry, the motion of random walk on the lines will be the same as if there was a single line. Therefore we can replace them with a single line but with different probability of hopping to line at the origin. That is to say $p_l=n_l/(n_l+2)$ and $p_p=2/(n_l+2)$.
We now ask the following question: could the asymptotic behavior, which we obtained above, be changed in favor of the line by increasing the number of lines $n_l$? The answer is no, because changing $p_l$ does not change the $z$ dependence of $P_{\textbf{0}}(z)$ at the limit $z\rightarrow 1$ and therefore the asymptotic behavior will be dominated by the plane. However the probability $p_b$ will set a time scale before which the behavior is effectively one-dimensional and then crosses over to two-dimensional. The time scale tends to infinity as the probability $p_l$ tends to one. We define the cross over time $t^*=-1/\ln z^*\approx 1/(1-z^*)$ where the $z^*$ is the value at which two terms in the brackets in Eq. (\ref{P0}) become of the same order, $\frac{p_p}{P^p_\textbf{0}(z^*)}=\frac{p_l}{P^l_\textbf{0}(z^*)}$. For small $p_p$ this condition is fulfilled at a value of $z^*$ very close to one. At this limit we use the approximate forms of these probabilities (\ref{P0-l}),(\ref{P0-p}) which gives $\frac{1}{\pi}\sqrt{2(1-z^*)}\ln[1/(1-z^*)]= p_p/p_l$. This is a transcendental equation for $z^*$ and thus the solution is not algebraic, however, it can be easily shown that $1-z^*$ approaches zero faster than $(p_p/p_l)^2$ and slower than $(p_p/p_l)^{(2+\delta)}$ for any positive $\delta$.

\begin{figure}
\centering
\includegraphics[width=0.47\textwidth]{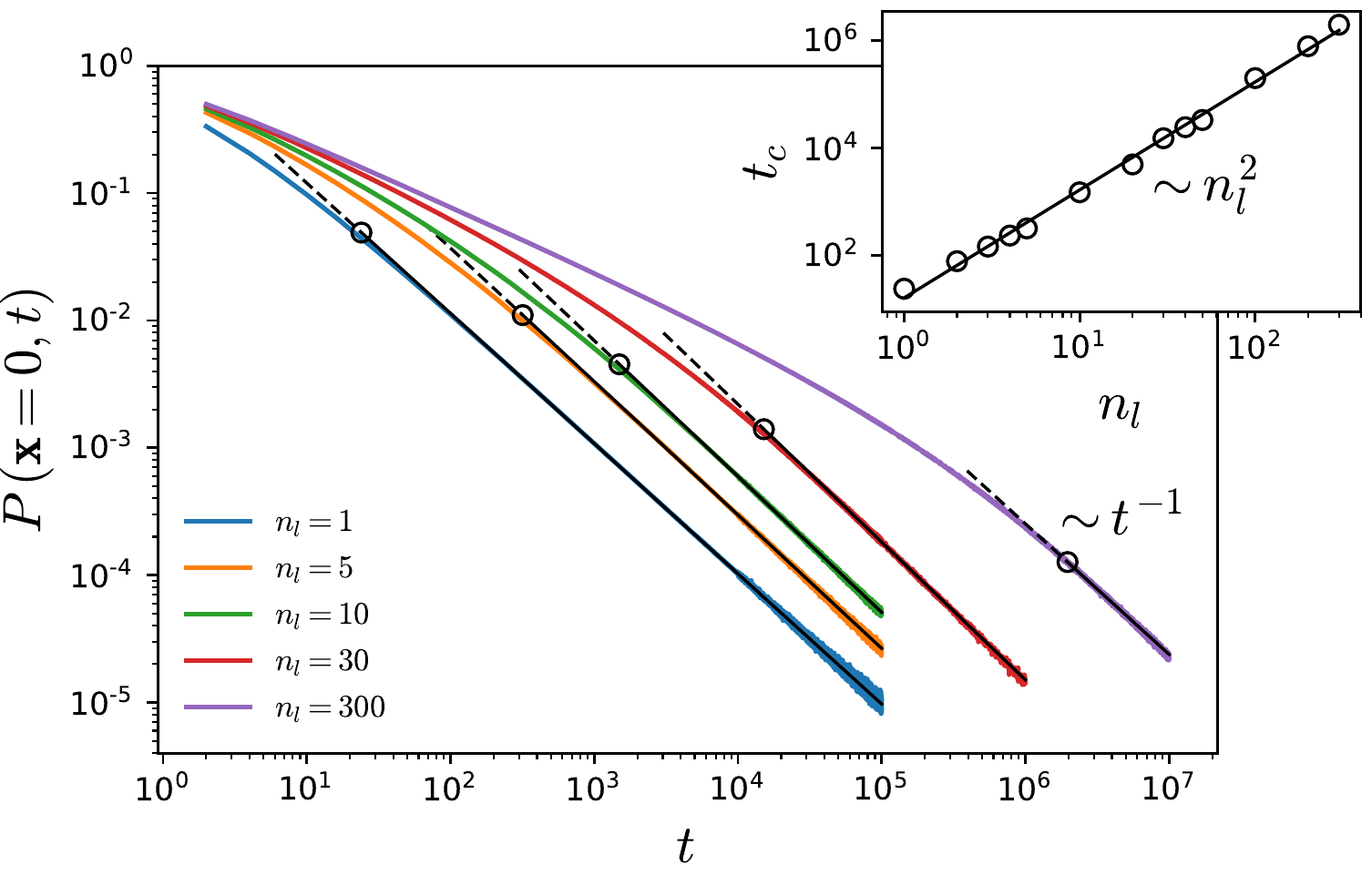}
\caption{(color online) Main: The probability to find the random walker at the origin $\textbf{x}=0$ at time $t$, i.e., $P(\textbf{x}=0, t)$, as function of $t$ in logarithmic scale for different number of crossing lines $n_l=1$ to $300$ from left to right. The open circles mark the crossover time $t_c$ after which the behavior is governed by the plane geometry with known scaling relation $\sim t^{-1}$ shown by the solid lines fitted to our data and followed by the dashed lines before $t_c$ when the behavior is still affected by the crossing line geometries. Inset: The crossover time $t_c$ as a function of the number of crossing lines $n_l$. The solid line is the best power-law fit to our data $t_c\sim n_l^2$ in perfect agreement with our analytical prediction---see the text.    }
\label{fig:Prob}
\end{figure}

\section{$1d$ Biased walk and Square lattice}
\label{biased}
A simple generalization of previous case which turns out to be important is to
consider asymmetric walk on the line. We denote different jump probabilities to the
right by $p$, and to the left by $q=1-p$. This also can be a representation of the walk on the Bethe lattice \cite{Sahimi1982,Cassi1989,Monthus1996}.  We should only replace the generating
function of site occupation probabilities of the line with
\begin{eqnarray}\label{mu0l}
 P^l_{\textbf{x}}(z)&=&(1-4pq z^2)^{-1/2}(1-\sqrt{1-4pqz^2})^{|x|} \nonumber\\ &&\times \left\{
   \begin{array}{cc}
    (2qz)^{-|x|} & x>0 \\
    (2pz)^{-|x|} & x<0.
   \end{array}\right.
\end{eqnarray}
It can be shown that the biased walk on the chain is not recurrent. More quantitatively $R^l=1-|2p-1|$ which is less than one if $p\neq\frac{1}{2}$. As a result, the RW on the combined geometry will no longer be recurrent as we have $R=1-\frac{1}{3}|2p-1|$.
In contrast to the previous case, the occupation of the origin  is dominated by the behavior of line. We can see that  $P_{\textbf{0}}^l(z)$ is convergent in the limit $z\rightarrow 1^-$. As a result, Eq. (\ref{P0}) gives $P_\textbf{0}(z)\approx \frac{1}{3}P^l_\textbf{0}(z)$. Now it can easily be shown that $P_\textbf{0}(2n)\sim \frac{1}{3} (\pi n)^{-1/2}(4pq)^n$. It is also interesting to note that the first moment i.e. the probability of finding the walker on the line is approaching one which means that even an infinitesimal amount of drift on the line will pull the walker into the line.

\section{numerical simulations}\label{simul}
In this section we present the results of our extensive numerical simulations of the model discussed in the previous sections and compare them with our analytical predictions.
We consider systems of combined geometries composed of a lattice plane and various number $n_l$ of lattice lines $ n_l = 1, 2, 3, 4, 5, 10, 20, 30, 40, 50, 100, 200, 300 $. The total simulation time for the cases $ n_l =1, 2,3,4,5,10 $ and $ 20,30,40,50 $ and $ 100, 200, 300 $ are taken to be $ 10^5, 10^6, 10^7 $, respectively, to be able to capture their corresponding asymptotic behavior. All measured quantities are averaged over more than $ 5\times 10^8 $ independent samples for each case. We assume that the random walker starts moving from origin at $t_0=0$ in all computations.

The first natural and standard quantity of interest is the mean squared displacement (MSD) of the RWs on the combined geometries. In order to see the individual contribution of the lattice plane and the lines, we have computed MSD for the plane (i.e., $\langle \textbf{r}^2_p\rangle$) and the lines(i.e., $\langle z_l^2\rangle$) at time $t$ separately. Figures \ref{fig:rp2} and \ref{fig:z2} show the corresponding dynamical evolution of MSD for various number of crossing lines. As shown in Fig. \ref{fig:rp2}, for larger number of crossing lines at the beginning times, the probability for the walker to go to one of the line geometries is higher ($n_l/(2+n_l)$) than the plane (see also Fig. \ref{fig:Pl}) and therefore, if the walker goes to the plane from origin (with probability $2/(2+n_l)$) it is attracted by the lines back to the origin which leads to the decrease in MSD on the plane for $n_l\gg 1$. At very large times, instead, the plane geometry will become dominant and the asymptotic behavior of the walker converges to a normal diffusion on a plane with the diffusion constant $D_p=1$. This explains why all plots for different $n_l$ converge to the same asymptotic value in Fig. \ref{fig:rp2}.

Figure \ref{fig:z2} also presents MSD on a line for various $n_l$. The dashed-lines show our analytical predictions for each $n_l$ for the MSD on a line at the very long-time limit which is $n_l$-dependent, and shows perfect agreement between our numerical simulations and analytical approximations. Notice that, unlike the symmetric diffusion on a single line geometry (which is known to behave as $\langle z_l^2\rangle\sim t$), the asymptotic behavior of the walks on the lines in our model does not follow the free diffusion and is governed by the square root of time containing a logarithmic correction, i.e, $\langle z_l^2\rangle\sim \sqrt{2/\pi^3}  n_l \sqrt{t} \log t$ for $t \gg 1$.

To better quantify the competition between the plane geometry and the crossing lines, we have computed the probability of finding the random walker at the origin $\textbf{x}=0$ at time $t$, i.e., $P(\textbf{x}=0, t)$, as function of $t$ for various $n_l$. As shown in Fig.  \ref{fig:Prob}, the behavior of $P(\textbf{x}=0, t)$ shows two primary and asymptotic regimes roughly for $t<t_c$ and $t>t_c$, respectively, for a $n_l$-dependent crossover time $t_c$. We find that at early times $t\ll t_c$ the behavior is governed by $\sim t^{-1/2}$ for the line geometry and crosses over to the long-time limit $t\ge t_c$ with $\sim t^{-1}$ for the plane geometry. For every $n_l$, we define $t_c$ as the (approximately) first time after which the scaling behavior of $P(\textbf{x}=0, t)$ is given by $\sim t^{-1}$ (marked by the open circle symbols in the Main Fig. \ref{fig:Prob}). The Inset of Fig. \ref{fig:Prob} shows the scaling relation between the crossing time and the number of crossing lines as $t_c\sim n_l^2$, which is in close agreement with our analytical approximations discussed at the end of Section \ref{example}.

\bigskip

\section{conclusions}\label{concl}
We have studied analytically the random walks problem on a combined lattice geometry composed of two generalized lattices with a single common point. After a general formulation of the problem, we illustrated the consequences in some nontrivial interesting examples by considering a lattice plane crossed by $n_l$ number of lattice lines at the origin. We have found that the probability of returning to the starting point at a long time limit is governed by the plane. The total probability of being in the line geometry increases first at the beginning time and then starts to decrease at larger times. Mean squared displacement asymptotically converges to the normal diffusion on the plane but it behaves like $\sqrt{t}\ln t$ on the line geometries. We have shown that the crossover time from the primary to the asymptotic regimes, scales approximately as $t_c\sim n_l^2$. Rather simple corollary is that the walk will be recurrent if it is recurrent on both lattices and will be transient if it is transient on at least one of them. We also examined the stability of the asymptotic behavior of the walk by introducing a perturbation to the model with a drift term along the line geometry (for $n_l=1$). We have found that even an infinitesimal amount of drift can totally change the asymptotic behavior of the walk in a way that the line geometry will dominate the long-time behavior of the perturbed model.

Our problem can also be viewed as a normal diffusion on a lattice plane with a single defect-site of variable waiting time. In this context, there has been studied \cite{Majumdar2017} a random reset problem on a $d$-dimensional lattice containing one trapping site with an exponential waiting time at the defect which exhibits a localization-delocalization phase transition. In our case, however, the waiting time is a power-law given by the diffusion along the crossing lines tuned by their number.

\section*{Acknowledgements}

R.S. and A.A.S. acknowledge partial financial support from the research council of the University of Tehran. A.A.S. would like to also acknowledge support from the Alexander von Humboldt Foundation.


\end{document}